\documentclass[12pt]{article}
\begin{document}
%%%%%%%%%%%%%%%%%%%%%%%%%%%%%%%%%%%%%%%%%%%%%
\baselineskip 15pt

\centerline{\Large{\bf Invariant Varieties of Periodic Points}}
\centerline{\Large{\bf for Some Higher Dimensional Integrable Maps}}
\vglue1cm
\centerline{Satoru SAITO~$^\dag$ and Noriko SAITOH~$^\ddag$}

\centerline{$^\dag$~Hakusan 4-19-10, Midori-ku, Yokohama 226-0006, Japan} 

\centerline{email: saito@phys.metro-u.ac.jp} 

\centerline{$^\ddag$~Applied Mathematics, Yokohama National University,}

\centerline{Hodogaya-ku, Yokohama 240-8501, Japan}

\centerline{email: nsaitoh@ynu.ac.jp}
\vglue1cm
\begin{center}
\begin{minipage}{14cm}
\noindent
{\bf Abstract}
By studying various rational integrable maps on $\mathbf{\hat C}^d$ with $p$ invariants, we show that periodic points form an invariant variety of dimension $\ge p$ for each period, in contrast to the case of nonintegrable maps in which they are isolated. We prove the theorem: {\it `If there is an invariant variety of periodic points of some period, there is no set of isolated periodic points of other period in the map.'}
\end{minipage}
\end{center}
%%%%%%%%%%%%%%%%%%%%%%%%%%%%%%%%%%%%%%%
\section{Introduction}

When a map is given there is no immediate way to foresee the fate of its iterations or whether it is integrable or nonintegrable. If the map is integrable the effect of a small difference of initial values remains small, while it behaves chaotically for some initial values if the map is nonintegrable. In spite of this large difference one cannot easily distinguish the two cases.

It would be desirable to know a way to distinguish integrable maps from nonintegrable ones just by investigation of the first few steps. This article is a contribution to finding such a criterion in the case of some higher dimensional maps.

The notion of integrability of a map has been studied in various contexts, such as the inverse method, the B\"aklund transformation method, the Darboux transformation method, the direct method, the geometrical method, etc. They judge the integrability of a map by testing if the map belongs to the classes to which the methods are applicable. Since the procedure is rather difficult, they will not decide if the information is limited only to the first few steps of the map.

In the case of continuous time Hamilton systems the Liouville theorem guarantees integrability of the systems if there are sufficient number of involutive invariants. There is no such criterion if the time evolution is discrete. In fact, when $d-1$ invariants are given arbitrary, we can always derive a map which reduces, for example, to the logistic map after the elimination of $d-1$ variables by using the invariants. 

In the theory of ordinary 2nd order differential equations, nonexistence of movable branch point singularities, dependent on initial values, indicates integrability of the equation (the Painlev\'e test). The method of singularity confinement was proposed \cite{GRPH, GRPH2} to replace this test and apply to discrete maps, but it does not always work \cite{HV}. As far as one dimensional maps are concerned there have been intensive studies mainly from the mathematical point of view \cite{Devaney, FM, FM2, Beardon}. It is known, for example, that the appearance of a Julia set, the closure of the set of repulsive periodic points, characterizes nonintegrability of the map. On the other hand very little is known about higher dimensional maps. In practice the best we can do at present is to study statistical information of the maps \cite{HV, BV}.\\

We consider a rational map on $\mathbf{\hat C}^d$, where $\mathbf{\hat C}=\{\mathbf{C},\infty\}$,
\begin{equation}
\mathbf{x}=(x_1,x_2,...,x_d)\quad \rightarrow\quad
\mathbf{X}=(X_1,X_2,...,X_d)=:\mathbf{X}^{(1)}.
\label{x->X}
\end{equation}
Throughout this paper we count the degree of freedom $d$ only which are coupled and are dependent each other. We are interested in the behaviour of the sequence: 
$\mathbf{x}\rightarrow \mathbf{X}^{(1)}\rightarrow\mathbf{X}^{(2)}\rightarrow\cdots.$ In particular we pay attention, in this paper, to the behaviour of periodic points of rational maps. If the map is nonintegrable we may find a set of isolated points with fractal structure as a higher dimensional counterpart of the Julia set\cite{BS}. We would like to know what object appears when the map is integrable. \\

We study in \S 2 the general feature of the periodicity conditions ${\mathbf X}^{(n)}={\mathbf x}$ of a rational map with $p\ (\ge 0)$ invariants. After the elimination of $p$ variables there are $d-p$ such conditions for each period. If they are independent, we obtain a set of isolated periodic points in general. We call the periodicity conditions of this generic type `uncorrelated'. It may happen that the $d-p$ conditions are correlated each other. In this case we find varieties of periodic points instead of isolated points. If all of the $d-p$ conditions are correlated, {\it i.e.}, when the conditions are `fully correlated', the varieties will be determined only by the invariants. We call such a variety `an invariant variety of periodic points'. Their precise meaning will be explained in \S 2. We prove there the following theorem:\\

\noindent
{\bf Theorem}

{\it 
If there is an invariant variety of periodic points of some period, there is no set of isolated periodic points of other period in the map.
}\\

Although the theorem doesn't tell us directly whether the map is integrable or nonintegrable, it is remarkable that it discriminates clearly a particular type of maps with an invariant variety of periodic points from the rest. In generic case with no invariant variety the periodicity conditions are uncorrelated and the periodic points form, for each period, a set of isolated points. In fact, if a rational map is given arbitrary and we calculate the periodicity conditions of any period, we find them uncorrelated almost always. 

Now we recall that almost all nonlinear maps which are chosen arbitrary are nonintegrable in general. This means that the chance for a nonintegrable map not to have a set of uncorrelated periodicity conditions of \underline{any period} must be practically zero. A typical higher dimensional nonintegrable map has a higher dimensional analog of the Julia set. In this case the periodic points form a closure of fractal sets of isolated points which become the source of chaotic orbits.

The precise notion of nonintegrability of a map has not been known. This makes difficult to discriminate integrable maps from nonintegrable ones unambiguously and prevents us to proceed discussion further. In order to overcome this problem we are going to adopt the facts we discussed above as our working hypothesis:\\

{\it If a map is nonintegrable, there is a set of uncorrelated periodicity conditions of some period.}\\

Note that we do not require that the periodicity conditions of all periods of a nonintegrable map are uncorrelated. We also emphasize that the existence of uncorrelated periodicity conditions of some period does not mean nonintegrability. Once we adopt the hypothesis, the following statement will be justified directly from our theorem.\\

\noindent
{\bf Conjecture}:

{\it
If there is an invariant variety of periodic points of some period, the map is integrable.}\\

Note that our conjecture does not exclude the possibility that some integrable maps do not have an invariant variety of periodic points. For example an integrable map with no invariant does not have an invariant variety. On the other hand there are $d$ dimensional maps which have $d-1$ invariants but reduce to some nonintegrable maps after elimination of $d-1$ variables. Therefore the situation is quite different from the continuous time Hamiltonian flow, whose integrability is guaranteed by the Liouville theorem if there are sufficient number of invariants.\\

We are not going to prove above hypothesis in this paper. But it is not difficult to convince ourselves of the fact as we calculate the periodicity conditions of any period of a rational map which is chosen arbitrary. In order to support our conjecture we would like to discuss in \S 3 and \S 4 if a set of periodic points of any period forms an invariant variety when the map is known integrable and there are sufficient number of invariants. As far as we have examined, we have found that all periodicity conditions possess this property if the map is integrable and periodic points exist, while no such property has been found otherwise. We show in \S 3 some examples of integrable maps, such as the Lotka-Volterra maps and the Toda maps. We also study the $q$-Painlev\'e maps which are integrable but not volume preserving in general. We find invariant varieties only when the parameters are restricted to have invariants. 

In \S 4 we discuss one dimensional maps to which some of higher dimensional integrable maps, such as the symmetric QRT map, the 3d Lotka-Volterra map, the $q$-Painlev\'e IV map with certain coefficients, can be reduced by using invariants. Analysis of these maps enables us to derive iteratively infinite series of invariant varieties of all periods. 

The method developed in \S 4 enables us to study properties of many higher dimensional maps all together simply by studying a single one dimensional map. We apply this method in \S 5 to investigate the integrable-nonintegrable transition of higher dimensional maps. In particular we consider a simple one dimensional map to which many higher dimensional maps can be reduced and interpolates between the M\"obius map and the logistic map. By studying the behaviour of periodic points in detail we clarify how the Julia set collapses and the invariant varieties of periodic points are created when the parameter of interpolation approaches to the critical value of the transition. This example shows explicitly the phenomenon which takes place at the border between integrable and nonintegrable regimes of higher dimensional maps.

%%%%%%%%%%%%%%%%%%%%%%%%%%%%%%%%%%%%%%%%%%%%%%%%%
\section{Nature of the Periodicity Conditions}

We study in this section the nature of the periodicity conditions of a rational map on $\mathbf{\hat C}^d$ which has $p\ (\ge 0)$ invariants. The periodic points of period $n$ will be found by solving
\begin{equation}
X_j^{(n)}=x_j,\qquad j=1,2,...,d.
\label{periodic conds}
\end{equation}
If $H_1(\mathbf{x}),H_2(\mathbf{x}),\cdots,H_p(\mathbf{x})$ are the invariants, the solutions of (\ref{periodic conds}) are constrained on an algebraic variety of dimension $d-p$ specified by the set of equations
\begin{equation}
H_i(\mathbf{x})=h_i,\qquad i=1,2,...,p.
\end{equation}
Here $h_1,h_2,...,h_p$ are the values of the invariants determined by the initial point of the map. Let us denote this variety by $V(h)$, {\it i.e.},
\begin{equation}
V(h)=\Big\{\mathbf{x}\Big|\ H_i(\mathbf{x})=h_i,\ i=1,2,...,p\Big\}.
\label{V(h)}
\end{equation}

The problem of finding periodic points is equivalent to finding an ideal generated by the set of $d+p$ functions $\{X_j^{(n)}(\mathbf{x})-x_j,\ \ H_i(\mathbf{x})-h_i\}$. Since the existence of the invariants enables us to eliminate $p$ components of $\mathbf{x}$ from (\ref{periodic conds}), the ideal reduces to the $p$th elimination ideal generated by certain functions $\Gamma^{(n)}_\alpha$ satisfying
\begin{equation}
\Gamma^{(n)}_\alpha(h_1,h_2,...,h_p,\xi_1,\xi_2,...,\xi_{d-p})=0,\qquad \alpha=1,2,...,d-p,\quad n\ge 2.
\label{Gamma_n=0}
\end{equation}
Here by $\xi_1,\xi_2,...,\xi_{d-p}$ we denote the variables which parameterize the variety $V(h)$ after the elimination of the $p$ components of $\mathbf{x}$. Note that the fixed point conditions $(n=1)$ are excluded in (\ref{Gamma_n=0}) since they are nothing to do with the invariants.

For an arbitrary set of values of $h_1,h_2,...,h_p$, the functions $\Gamma_\alpha^{(n)}(h,\mathbf{\xi})$ define an affine variety, which we denote by $V^{(n)}(\langle \Gamma\rangle)$, {\it i.e.},
\begin{eqnarray*}
V^{(n)}(\langle\Gamma\rangle)=\Big\{\mathbf{\xi}\Big|\ \Gamma^{(n)}_\alpha(h,\mathbf{\xi})=0,\ \ \alpha=1,2,...,d-p\Big\},\quad n\ge 2.
\end{eqnarray*}
In general this variety consists of a finite number of isolated points on $V(h)$, hence zero dimension, corresponding to the solutions to the $d-p$ algebraic equations (\ref{Gamma_n=0}) for the $d-p$ variables $\xi_1,\xi_2,...,\xi_{d-p}$. Once the values of $\xi_1,\xi_2,...,\xi_{d-p}$ are decided by solving (\ref{Gamma_n=0}), the location of a periodic point on $\mathbf{\hat C}^d$ will be determined from the information of the values of $h_1,h_2,...,h_p$. In this case we obtain a number of isolated periodic points of period $n$, and say that the periodicity conditions (\ref{periodic conds}) are `uncorrelated'. Needless to say this case includes a map with no invariants. \\

There are possibilities that the equations $(\ref{Gamma_n=0})$ impose relations on $h_1,h_2,...,h_p$ instead of fixing all $\xi_\alpha$'s. Let $l$ be the number of such equations. We write them as
\begin{equation}
\gamma^{(n)}_\alpha(h_1,h_2,...,h_p)=0,\qquad \alpha=1,2,...,l,
\label{gamma_n}
\end{equation}
instead of $\Gamma^{(n)}_\alpha$, to emphasize independence from $\xi_j$'s. If $m$ is the number of the rest of the equations
\begin{equation}
\Gamma^{(n)}_\alpha(h_1,h_2,...,h_p,\xi_1,\xi_2,...,\xi_{d-p})=0,\qquad \alpha=1,2,...,m
\label{Gamma_n}
\end{equation}
$d-p-m$ variables are not determined from the periodicity conditions. This means that $V^{(n)}(\langle\Gamma\rangle)$ forms a subvariety of dimension $d-p-m$ of $V(h)$.
We say that the periodicity conditions are `correlated' in this case. Since the conditions (\ref{gamma_n}) leave $p-l$ invariants free, there remain totally $d-m-l$ variables undetermined by the periodicity conditions. In other words the points of period $n$ satisfying the conditions (\ref{gamma_n}) and (\ref{Gamma_n}) form a subvariety of dimension $d-m-l$ in $\hat{\mathbf C}^d$.

When $m=0$ the periodicity conditions determine none of the variables $\xi_1,\xi_2,...,\xi_{d-p}$ but impose $l$ relations among the invariants. Then the affine variety $V^{(n)}(\langle\Gamma\rangle)$ coincides with $V(h)$. In other words every point on $V(h)$ is a periodic point of period $n$, while $V(h)$ itself is constrained by the relations among the invariants. We say that the periodicity conditions are `fully correlated' in this particular case. If we replace $h_i$ by $H_i({\mathbf x})$ in $\gamma^{(n)}_\alpha(h)$ the periodicity conditions
\begin{equation}
\gamma^{(n)}_\alpha(H_1(\mathbf{x}),H_2(\mathbf{x}),...,H_p(\mathbf{x}))=0,\qquad \alpha=1,2,...,l.
\label{gamma=0}
\end{equation}
enable us to consider the constraints on the invariants as constraints on the variables $\mathbf{x}$. We denote by $v^{(n)}(\langle \gamma\rangle)$ the affine variety generated by the functions $\gamma^{(n)}_\alpha(H_1(\mathbf{x}),H_2(\mathbf{x}),...,H_p(\mathbf{x}))$, and distinguish it from $V^{(n)}(\langle \Gamma\rangle)$. Namely we define
\begin{eqnarray*}
v^{(n)}(\langle\gamma\rangle)=\Big\{\mathbf{x}\Big|\ \gamma^{(n)}_\alpha(H_1(\mathbf{x}),H_2(\mathbf{x}),...,H_p(\mathbf{x}))=0,\ \ \alpha=1,2,...,l\Big\}.
\end{eqnarray*}
\ \\

The significance of defining $v^{(n)}(\langle \gamma\rangle)$ lies in the fact that if a point belongs to $v^{(n)}(\langle \gamma\rangle)$ the point is a periodic point of period $n$. This is true only in the case $m=0$. If $m>0$, the constraints on the invariants are not sufficient to determine the periodic points. We also notice that the variety $v^{(n)}(\langle \gamma\rangle)$ is determined by the invariants $h_1,h_2,...,h_p$ alone as is clear from the construction. We have called $v^{(n)}(\langle \gamma\rangle)$ `an invariant variety of periodic points' in \S 1. Let us summarize the properties of $v^{(n)}(\langle \gamma\rangle)$ as follows.
\begin{itemize}
\item
The dimension of $v^{(n)}(\langle \gamma\rangle)$ is $d-l\ (\ge p)$.
\item
Every point on $v^{(n)}(\langle \gamma\rangle)$ can be an initial point of the periodic map of period $n$.
\item
All images of the periodic map starting from a point of $v^{(n)}(\langle \gamma\rangle)$ remain on it.
\item
$v^{(n)}(\langle \gamma\rangle)$ is determined by the invariants of the map alone.
\end{itemize}
\vglue0.5cm

Having introduced some notions about the periodicity conditions we are now going to establish our theorem presented in \S 1. For this purpose let us suppose that the periodicity conditions of period $n$ are given by (\ref{gamma_n}) and (\ref{Gamma_n}). They are fully correlated if $m=0$, and correlated if $d-p>m\ge 0$. Note that, since the $l$ relations among the invariants are assumed independent, the conditions (\ref{gamma_n}) are consistent only if
\begin{equation}
p\ge l,\qquad d-p\ge l+m
\label{p ge l}
\end{equation}
are satisfied. We also emphasize that the conditions (\ref{gamma_n}) include neither fixed points nor points whose periods are the divisors of $n$. \\

We further assume that the periodicity conditions of period $k(\ne n)$ are uncorrelated, {\it i.e.},
\begin{equation}
\Gamma^{(k)}_\alpha(h_1,h_2,...,h_p,\xi_1,\xi_2,...,\xi_{d-p})=0,\qquad \alpha=1,2,...,d-p,\qquad k\ne n
\label{Gamma_k}
\end{equation}
and ask if there exist $\xi_1,\xi_2,...,\xi_{d-p}$ which satisfy (\ref{gamma_n}), (\ref{Gamma_n}) and (\ref{Gamma_k}) simultaneously for some values of the invariants $h_1,h_2,...,h_p$. It is necessary, but not sufficeint, for this to happen that the total number $d$ of the variables $\xi_1,\xi_2,...,\xi_{d-p}$ and the invariants $h_1,...,h_p$ exceeds the total number $d-p+l+m$ of the equations (\ref{gamma_n}), (\ref{Gamma_n}) and (\ref{Gamma_k}), {\it i.e.}, $p\ge l+m$ must hold. Combining this together with (\ref{p ge l}) we obtain
\begin{equation}
\min\{p,d-p\}\ge l+m.
\label{min p,d-p ge l+m}
\end{equation}

Now we suppose that (\ref{min p,d-p ge l+m}) is satisfied and (\ref{Gamma_k}) and (\ref{Gamma_n}) are solved to determine all $\xi$'s and $m$ invariants, say $h_1,h_2,...,h_{m}$. Since the rest of the invariants, $h_{m+1},...,h_p$, are completely free it is aways possible to fix them such that (\ref{gamma_n}) is satisfied. Therefore it is inevitable, as far as $l\ne 0$ and the periodicity conditions of period $k$ is uncorrelated, there exist some points which have two different periods $n$ and $k$ simultaneously. This is certainly a contradiction. Thus we are lead to the following conclusion:\\ 

\noindent
{\bf Lemma}

{\it A set of correlated periodicity conditions satisfying (\ref{min p,d-p ge l+m}) and a set of uncorrelated periodicity conditions of a different period do not exist in one map simultaneously.}\\

In particular when the periodicity conditions of period $n$ are fully correlated, {\it i.e.}, when $m=0$ in (\ref{Gamma_n}), the condition (\ref{min p,d-p ge l+m}) is always satisfied as long as (\ref{gamma_n}) has solutions. Our theorem in \S 1 follows from this fact immediately.\\

Some remarks are in order:

\begin{enumerate}
\item
The existence of an invariant variety of periodic points is a sufficient, but not necessary, condition for the integrablity. 
\item
The existence of a set of isolated periodic points is a necessary, but not sufficient, condition for the nonintegrability.

\item
When $p<d/2$, a set of correlated periodicity conditions satisfying $d-p\ge l+m>p\ge l\ge 0$ and a set of uncorrelated conditions of other period can exist together in one map. 
\item
When $p\ge d/2$, uncorrelated periodicity conditions of two different periods can exist together, although the total number of the variables exceeds the number of the conditions. This is because the argument, which lead us to the Lemma, does not apply when $l=0$. Therefore there are possibilities that all periodicity conditions are uncorrelated even though the map has invariants more than $d/2$. A simple example will be discussed in \S 4. 
\end{enumerate}

%%%%%%%%%%%%%%%%%%%%%%%%%%%%%%%%%%%%%%%%%%%%%%%%%%%%%%%%%%%%%%%%%%%%%%%%%%%%%%%%%%%%%%%%%%%%%%%%%%%%%%%%

\section{Invariant Varieties of Periodic Points}

In order to see the correspondence between the existence of an invariant variety of periodic points and integrability of the map, we are going to study in this section various integrable maps and see if the periodicity conditions (\ref{periodic conds}) exhibit the invariant varieties of periodic points.

Since we need assistance of computer, the procedure of finding the invariant varieties of many variables and/or of higher degree becomes harder as the dimension of the map and/or the degree of period increases. Nevertheless, as far as we are able to calculate, we are convinced that all perodic points, if they exist, satisfy fully correlated periodicity conditions and form an invariant variety for each period, as we present below.

%%%%%%%%%%%%%%%%%%%%%%%
\subsection{Lotka-Volterra maps}

Although there have been intensive studies of discrete integrable systems \cite{Hirota, QRT, QRT2, Suris, Suris2}, the precise notion of integrability of a map has not been known. Among others the bilinear method provides a powerful tool to find integrable maps, in the sense that the solutions are given explicitly in terms of the $\tau$ function of the KP hierarchy \cite{Hirota, HTI, HTI2}. We study, in this section, the Lotka-Volterra maps, the Toda map, and the $q$-Painlev\'e maps, all derived by this method. 

The Lotka-Volterra map $\mathbf{x}\rightarrow\mathbf{X}$ of dimension $d$ can be found \cite{HTI, HTI2} by solving
\begin{equation}
X_j(1-X_{j-1})=x_j(1-x_{j+1}),\qquad j=1,2,...,d
\label{LV eq}
\end{equation}
for $\mathbf{X}=(X_1,X_2,...,X_d)$ under the conditions $x_{j+d}=x_j\ (j=1,2,...,d)$. We show in the Appendix that the invariants of this map are given by
\begin{equation}
\left\{\begin{array}{cl}
H_k&={\sum}'_{j_1,j_2,...,j_k}x_{j_1}x_{j_2}\cdots x_{j_k}(1-x_{j_1-1})(1-x_{j_2-1})\cdots (1-x_{j_k-1})\cr
&\qquad\qquad\qquad\qquad ( k=1,2,...,[d/2] )\cr
r&=x_1x_2\cdots x_d\cr
\end{array}\right.
\label{H_k}
\end{equation}
Here the prime in the summation $\sum'$ of (\ref{H_k}) means that the summation must be taken over all possible combinations $j_1,j_2,...,j_k$ but excluding direct neighbours. The total number of the invariants is $p=[d/2]+1$, where $[d/2]=d/2$ if $d$ is even and $[d/2]=(d-1)/2$ if $d$ is odd. 

\subsubsection{3d and 4d Lotka-Volterra maps}

When $d=3$ the explicit form of the map is obtained as
\begin{equation}
X_1=x_1{1-x_2+x_2x_3\over 1-x_3+x_3x_1},\quad X_2=x_2{1-x_3+x_3x_1\over 1-x_1+x_1x_2},\quad X_3=x_3{1-x_1+x_1x_2\over 1-x_2+x_2x_3}.
\label{3LV}
\end{equation}
by solving (\ref{LV eq}) for $(X_1,X_2,X_3)$. There are two invariants
\begin{equation}
r=x_1x_2x_3,\qquad s=(1-x_1)(1-x_2)(1-x_3),
\label{r,s}
\end{equation}
where we used $s:=H_1/r$ instead of $H_1$ for convenience. Since $d-p=1$, we expect one invariant variety of periodic points for each period. After some manipulation we find the following results for the variety 
\begin{equation}
v^{(2)}(\langle \gamma\rangle)=\{\mathbf{x} |\ s+1=0\}
\label{period 2}
\end{equation}
for the period 2 case,
\begin{equation}
v^{(3)}(\langle \gamma\rangle)=\{\mathbf{x} |\ r^2+s^2-rs+r+s+1=0\}
\label{period 3}
\end{equation}
for the period 3 case,
\begin{equation}
v^{(4)}(\langle \gamma\rangle)=\{\mathbf{x} |\ r^3s+s^3-3rs^2+6r^2s+3rs-r^3+s=0\}
\label{period 4}
\end{equation}
for the period 4 case and
\begin{eqnarray}
v^{(5)}(\langle \gamma\rangle)&=&\{\mathbf{x} |\ r^3s^4-r^3s^2-6r^4s^5+10r^3s^6+3s^5r+s^6+s^5+3r^4s^4-3r^5s^3\nonumber\\
&&
-6r^4s^3-r^6s^3+3r^5s^4+s^4+21s^4r^2+6s^4r+r^3s^7+s^7+27s^5r^2\nonumber\\
&&
-3s^6r-r^3s^5+21r^2s^6-10r^3s^3-6rs^7+s^8=0\},
\label{period 5}
\end{eqnarray}
for the period 5 case, etc.. \\

The 4 dimensional Lotka-Volterra map is defined by
\begin{eqnarray*}
X_1=x_1{1-x_2-x_3+x_2x_3+x_3x_4\over 1-x_3-x_4+x_3x_4+x_4x_1},&\quad&
X_2=x_2{1-x_3-x_4+x_3x_4+x_4x_1\over 1-x_4-x_1+x_4x_1+x_1x_2},\nonumber\\
X_3=x_3{1-x_4-x_1+x_4x_1+x_1x_2\over 1-x_1-x_2+x_1x_2+x_2x_3},&\quad&
X_4=x_4{1-x_1-x_2+x_1x_2+x_2x_3\over 1-x_2-x_3+x_2x_3+x_3x_4}
.
\label{4dLV}
\end{eqnarray*}
There are three invariants of the map, given by
$$
r=x_1x_2x_3x_4,\qquad H_1=x_1+x_2+x_3+x_4-x_1x_4-x_2x_1-x_3x_2-x_4x_3,
$$$$
H_2=x_1x_3+x_2x_4-x_2x_3x_4-x_3x_4x_1-x_4x_1x_2-x_1x_2x_3+2x_1x_2x_3x_4.
$$
We find an invariant variety of periodic points
$$
v^{(2)}(\langle \gamma\rangle)=\{\mathbf{x} |\ H_1-2=0\}
$$
for the period 2 case, and
$$
v^{(3)}(\langle \gamma\rangle)=\{\mathbf{x} |\ H_2^2-H_1^2H_2+2H_1H_2-2rH_2-6H_2+H_1^3-2H_1^2$$$$+2rH_1+2H_1+r^2+2r+5=0\}
$$
for the period 3 case, etc..

%%%%%%%%%%%%%%%%%%%%%%%%%%%%%%%
\subsubsection{5d Lotka-Volterra map}

When $d=5$ there are three invariants, thus we have $d-p=2$. From our argument of \S 2 the periodicity conditions (\ref{periodic conds}) will be satisfied by imposing two independent conditions $\gamma_1^{(n)}=0,\ \gamma_2^{(n)}=0$, simultaneously, for each period. The situation changes from the previous cases $d=3,4$ where $d-p$ is 1.

By solving (\ref{LV eq}) in the case $d=5$ we obtain
{\small
\begin{eqnarray*}
X_1 &=& x_1{1-x_2-x_3-x_4+x_2x_3+x_2x_4+x_3x_4 +x_4x_5 -x_2x_3x_4-x_2x_4x_5 +x_2x_3x_4x_5
\over 
1-x_3-x_4-x_5+x_3x_4+x_3x_5+x_4x_5+x_5x_1-x_3x_4x_5-x_3x_5x_1+x_3x_4x_5x_1
}\\
X_2 &=& x_2{1-x_3-x_4-x_5+x_3x_4+x_3x_5+x_4x_5+x_5x_1-x_3x_4x_5-x_3x_5x_1+x_3x_4x_5x_1\over
1-x_4-x_5-x_1+x_4x_5+x_4x_1+x_5x_1+x_1x_2-x_4x_5x_1-x_4x_1x_2+x_4x_5x_1x_2}\\
X_3 &=& x_3{1-x_4-x_5-x_1+x_4x_5+x_4x_1+x_5x_1+x_1x_2-x_4x_5x_1-x_4x_1x_2+x_4x_5x_1x_2\over
1-x_5-x_1-x_2+x_5x_1+x_5x_2+x_1x_2+x_2x_3-x_5x_1x_2-x_5x_2x_3+x_5x_1x_2x_3
}\\
X_4 &=& x_4{1-x_5-x_1-x_2+x_5x_1+x_5x_2+x_1x_2+x_2x_3-x_5x_1x_2-x_5x_2x_3+x_5x_1x_2x_3\over 
1-x_1-x_2-x_3+x_1x_2+x_1x_3+x_2x_3+x_3x_4-x_1x_2x_3-x_1x_3x_4+x_1x_2x_3x_4}\\
X_5 &=& x_5{1-x_1-x_2-x_3+x_1x_2+x_1x_3+x_2x_3+x_3x_4-x_1x_2x_3-x_1x_3x_4+x_1x_2x_3x_4
\over 1-x_2-x_3-x_4+x_2x_3+x_2x_4+x_3x_4 +x_4x_5 -x_2x_3x_4-x_2x_4x_5 +x_2x_3x_4x_5}.
\end{eqnarray*}
}
From our general formula the three invariants are
\begin{eqnarray}
H_1&=&x_1x_2+x_2x_3+x_3x_4+x_4x_5+x_5x_1-x_1-x_2-x_3-x_4-x_5,\nonumber\\
H_2&=&x_1x_3+x_2x_4+x_3x_5+x_4x_1+x_5x_2
-x_1x_2x_3-x_2x_3x_4-x_3x_4x_5\nonumber\\
&&-x_4x_5x_1-x_5x_1x_2-x_1x_2x_4-x_1x_3x_4-x_1x_3x_5-x_2x_3x_5-x_2x_4x_5\nonumber\\
&&+x_2x_3x_4x_5+x_3x_4x_5x_1+x_4x_5x_1x_2+x_5x_1x_2x_3+x_1x_2x_3x_4,
\label{H_2 of d=5}\\
r&=&x_1x_2x_3x_4x_5.
\nonumber
\end{eqnarray}
If we form two particular combinations $H_2+3H_1+5$ and $H_1+r+2$ from these invariants, we see that the Gr\"obner basis of this ideal generates the 3rd elimination ideal of the functions $\{X_j^{(2)}-x_j\}$. Therefore the invariant variety of periodic points of period 2 is
$$
v^{(2)}(\langle \gamma\rangle)=\{\mathbf{x}|\ H_2+3H_1+5=0,\ H_1+r+2=0\}.
$$
This is an algebraic variety of dimension 3.

%%%%%%%%%%%%%%%%%%%%%%%%%%%%%%%%%%%%%%%%%%%%%%

\subsection{3 point Toda map}

The 3 point Toda map \cite{HTI, HTI2}
$$
(i_1,i_2,i_3,v_1,v_2,v_3)\ \ \rightarrow\ \ (I_1,I_2,I_3,V_1,V_2,V_3),
$$
is defined by
\begin{eqnarray*}
I_1 = i_2{i_3v_1+i_3i_1+v_3v_1\over i_2v_3+i_2i_3+v_2v_3},
&\quad&
V_1 = v_1{i_2v_3+i_2i_3+v_2v_3\over i_3v_1+i_3i_1+v_3v_1},\\I_2 = i_3{i_1v_2+i_1i_2+v_1v_2\over i_3v_1+i_3i_1+v_3v_1},
&\quad&
V_2 = v_2{i_3v_1+i_3i_1+v_3v_1\over i_1v_2+i_1i_2+v_1v_2},\\I_3 = i_1{i_2v_3+i_2i_3+v_2v_3\over i_1v_2+i_1i_2+v_1v_2},
&\quad&
V_3 = v_3{i_1v_2+i_1i_2+v_1v_2\over i_2v_3+i_2i_3+v_2v_3}.
\end{eqnarray*}
This map has four invariants,
\begin{eqnarray*}
t_1&=&i_1+i_2+i_3+v_1+v_2+v_3,\\
t_2&=&i_1i_2+i_2i_3+i_3i_1+v_1v_2+v_2v_3+v_3v_1+i_1v_2+i_2v_3+i_3v_1,\\
t_3&=&i_1i_2i_3,\\
t'_3&=&v_1v_2v_3.
\end{eqnarray*}

From these data we find the Gr\"obner bases generating the 4th elimination ideal of $\{X_j^{(2)}-x_j\}$ and $\{X_j^{(3)}-x_j\}$. The periodicity conditions for the period 2 case are not fully correlated but only correlated. Therefore we have no invariant variety of periodic points. In the case of period 3 the conditions are fully correlated. We obtain an invariant variety of periodic points of dimension 4 as follows,
$$
v^{(3)}(\langle \gamma\rangle)=\{\mathbf{x}|\ t_1=0,\ t_2=0\}.
$$

%%%%%%%%%%%%%%%%%%%%%%%%%%%%%%%%%%
\subsection{$q$-Painlev\'e maps}

We examine in this subsection discrete analogues of the Painlev\'e equations which do not belong to the Lotka-Volterra series. Since they do not have invariants in generic cases their integrability is not obvious at all. As we choose the parameters in the map properly there appear invariants and our theorem guarantees integrability of the map. Therefore they provide examples of the maps in the border to which our theorem can apply to discriminate integrable maps from nonintegrable ones.

There have been proposed various types of the discrete Painlev\'e equations which preserve integrability \cite{GR, KNY}. Among others we study here the symmetric versions. By choosing the dependent variables properly the $q$-Painlev\'e IV map can be written in the symmetric form \cite{KNY}
\begin{eqnarray}
X_1&=&\alpha_1\alpha_2x_2{1-\alpha_3x_3+\alpha_3\alpha_1x_3x_1\over 1-\alpha_1x_1+\alpha_1\alpha_2x_1x_2},\nonumber\\
X_2&=&\alpha_2\alpha_3x_3{1-\alpha_1x_1+\alpha_1\alpha_2x_1x_2\over 1-\alpha_2x_2+\alpha_2\alpha_3x_2x_3},
\label{PIV}\\
X_3&=&\alpha_3\alpha_1x_1{1-\alpha_2x_2+\alpha_2\alpha_3x_2x_3\over 1-\alpha_3x_3+\alpha_3\alpha_1x_3x_1}.\nonumber
\end{eqnarray}

This map is integrable in the sense that the map admits B\"acklund transformations. It is also supported by the existence of some type of explicit solutions. Since the Jacobian of this map equals $\alpha_1\alpha_2\alpha_3$, the map does not preserve the volume in general. Under these circumstances the periodicity conditions are uncorrelated for all periods. Hence this map is an example of integrable maps which do not have an invariant variety of periodic points. 

As we impose the condition $\alpha_1\alpha_2\alpha_3=1$, the map has one invariant
\begin{equation}
r=x_1x_2x_3,
\label{Painleve r}
\end{equation}
hence $p=1$. Because $p<d/2$ in this case, correlated and uncorrelated periodicity conditions of different periods can exist in one map, as we noticed in Remark (3) of \S 2. If we solve the periodicity condisions of period 2, we find an invariant variety
\begin{equation}
v^{(2)}(\langle\gamma\rangle)=\{{\mathbf x} |\ r+1=0\ \}
\label{r+1=0}
\end{equation}
and also a curve in $\hat {\mathbf C}^3$
\begin{eqnarray}
&&\{{\mathbf x}\ |(1-\alpha_3x_3)x_2^2-\alpha_2^{-1}(1-x_3^2)x_2-(x_3-\alpha_3)x_3=0\nonumber\\
&&
\qquad\cap\  
(\alpha_1\alpha_2-x_3)(x_1x_2-x_3)-\alpha_1x_1(1-x_3^2)=0\ \}.
\label{curve}
\end{eqnarray}
The first solution $v^{(2)}$ of (\ref{r+1=0}) corresponds to the fully correlated periodicity conditions of the case $l=1,m=0$. According to our conjecture the existence of $v^{(2)}$ guarantees integrability of the map. The second solution, the curve of (\ref{curve}), fixes isolated points on $V(h)$, corresponding to the uncorrelated conditions of the case $l=0,m=2$, as we fix the value of $r$. 

The result of period 2 provides us useful information to explore the implication of our theorem. The theorem tells us that the existence of $v^{(2)}$ excludes uncorrelated periodicity conditions of other period in this map. Thus (\ref{curve}) is the only possible set of periodic points which can generate isolated periodic points for each value of $r$. On the other hand the existence of the solution (\ref{curve}) excludes invariant varieties of other periods. Therefore the periodicity conditions of all other periods must produce surfaces in $\hat {\mathbf C}^3$ corresponding to $l=0,m=1$. In fact we find the points of period 3 only on the surface defined by
\begin{eqnarray*}
&&\alpha_1^2(1+\alpha_2^2+\alpha_2^2\alpha_1^2)(r^2+1)+\alpha_1(1+\alpha_1^2+3\alpha_2^2\alpha_1^2)(x_2x_3+rx_1)\\
&&+\alpha_1^2\alpha_2(\alpha_1^2+\alpha_2^2\alpha_1^2+3)(x_3x_1+rx_2)+\alpha_1\alpha_2(1+\alpha_2^2\alpha_1^2+3\alpha_1^2)(x_1x_2+rx_3)\\
&&
-(1+\alpha_1^2+\alpha_2^2\alpha_1^2)\Big(\alpha_1(\alpha_2x_1x_2+x_2x_3+\alpha_1\alpha_2x_3x_1)+(1+\alpha_1^2+\alpha_2^2\alpha_1^2)\Big)r\\
&&-\alpha_1(1+\alpha_1^2+\alpha_2^2\alpha_1^2)(x_1+\alpha_1\alpha_2x_2+\alpha_2x_3)\\
&&+\alpha_1^2(x_1^2+\alpha_1^2\alpha_2^2x_2^2+\alpha_2^2x_3^2)+\alpha_1^2(\alpha_2^2x_1^2x_2^2+x_2^2x_3^2+\alpha_1^2\alpha_2^2x_1^2x_3^2)\\
&&-2\alpha_1^2\alpha_2(x_1^2x_2+\alpha_1x_2^2x_3+\alpha_1\alpha_2x_1x_2^2+\alpha_1x_1^2x_3+\alpha_1\alpha_2x_1x_3^2+x_2x_3^2)
=0.
\end{eqnarray*}
For each value of the invariant $r$ this surface generates a curve on $V(h)$. Similarly we have found a surface formed by points of period 4, but more complicated.
\\

If we further fix the parameters to $\alpha_1=\alpha_2=\alpha_3=1$, the set of equations (\ref{PIV}) is obtained from the 3d Lotka-Volterra map (\ref{3LV}) simply by the shift $(X_1,X_2,X_3)\rightarrow (X_3,X_1,X_2)$. This map has two invariants, one is $r$ of (\ref{Painleve r}) and the other one is $s=(1-x_1)(1-x_2)(1-x_3)$, the same as (\ref{r,s}). In this particular case we have found a series of the invariant varieties of periodic points
\begin{eqnarray}
v^{(2)}(\langle\gamma\rangle)&=&\{\mathbf{x}|\ r+1=0\}\nonumber\\
v^{(3)}(\langle\gamma\rangle)&=&\{\mathbf{x}|\ r^2+s^2-rs+r+s+1=0\}\label{PIVinvariant varieties}\\
v^{(4)}(\langle\gamma\rangle)&=&\{\mathbf{x}|\ s^3r+r^3+6rs^2+3rs-s^3+r-3r^2s=0\},\nonumber
\end{eqnarray}
and so on.\\

The symmetric version of the $q$-Painlev\'e V map has been given explicitly \cite{Masuda} by
\begin{eqnarray}
X_1&=&\alpha_1\alpha_2x_2{1-\alpha_3x_3+\alpha_3\alpha_4x_3x_4-\alpha_3\alpha_4\alpha_1x_3x_4x_1\over 1-\alpha_1x_1+\alpha_1\alpha_2x_1x_2-\alpha_1\alpha_2\alpha_3x_1x_2x_3},\nonumber\\
X_2&=&\alpha_2\alpha_3x_3{1-\alpha_4x_4+\alpha_4\alpha_1x_4x_1-\alpha_4\alpha_1\alpha_2x_4x_1x_2\over 1-\alpha_2x_2+\alpha_2\alpha_3x_2x_3-\alpha_2\alpha_3\alpha_4x_2x_3x_4},\nonumber\\
X_3&=&\alpha_3\alpha_4x_4{1-\alpha_1x_1+\alpha_1\alpha_2x_1x_2-\alpha_1\alpha_2\alpha_3x_1x_2x_3\over 1-\alpha_3x_3+\alpha_3\alpha_4x_3x_4-\alpha_3\alpha_4\alpha_1x_3x_4x_1},\nonumber\\
X_4&=&\alpha_4\alpha_1x_1{1-\alpha_2x_2+\alpha_2\alpha_3x_2x_3-\alpha_2\alpha_3\alpha_4x_2x_3x_4\over 1-\alpha_4x_4+\alpha_4\alpha_1x_4x_1-\alpha_4\alpha_1\alpha_2x_4x_1x_2}.
\label{Painleve V}
\end{eqnarray}
If we limit the parameters by $\alpha_1\alpha_2\alpha_3\alpha_4=1$ the map has one invariant
$$
r=x_1x_2x_3x_4,
$$
and an invariant variety of periodic points
\begin{equation}
v^{(2)}(\langle\gamma\rangle)=\{ {\mathbf x}|\ r-1=0\ \}
\label{r-1}
\end{equation}
in the period 2 case. If we further fix the parameters as $\alpha_1=\alpha_2=\alpha_3=\alpha_4=1$ there are two more invariants 
\begin{equation}
s=(1-x_1)(1-x_2)(1-x_3)(1-x_4),
\qquad 
t=(1-x_2x_4)(1-x_1x_3).
\label{invariants of Painleve}
\end{equation}
in addition to $r$, and (\ref{r-1}) is the invariant variety of periodic points of period 2 again.

As the period increases the computer manipulation becomes more difficult to derive the invariant varieties explicitly. For some cases we discussed in this section, the method we develop in \S 4 will enable us to find iteratively the invariant varieties of all periods if we know one of them.

%%%%%%%%%%%%%%%%%%%%%%%%%%%%%%%%%%%%%%%%%%%%%%%%%%
\section{Series of Invariant Varieties of Periodic Points}

We study in this section a method of generating an infinite series of invariant varieties of periodic points. Although application of this method is limited to some particular type of maps at present the existence of such method seems to suggest that the invariant varieties of different periods are correlated each other.

\subsection{Generation of higher dimensional maps}

Let us suppose that the $p$ invariants $H_1(\mathbf{x}), H_2(\mathbf{x}), ..., H_p(\mathbf{x})$ are given and we want to derive a $d$ dimensional map which reduces to the following  $d-p$ dimensional one
\begin{equation}
x_i\rightarrow X_i=f_i(x_1,x_2,...,x_{d-p}),\qquad i=1,2,...,d-p,
\label{X_i=f_i}
\end{equation}
after the elimination of the variables $x_{d-p+1},...,x_d$.
This will be done easily. If 
$$
X_i(\mathbf{x})=g_i(\mathbf{x}),\qquad i=d-p+1,...,d
$$
are the solutions of
$$
H_i(f_1,f_2,...,f_{d-p},X_{d-p+1},...,X_d)=H_i(\mathbf{x}),\quad i=1,2,...,p,
$$
then 
$$
\mathbf{x}\rightarrow \mathbf{X}=(f_1,...,f_{d-p},g_{d-p+1},...,g_d)
$$
is the $d$ dimensional map which satisfy our requirements. In this way we can derive many higher dimensional maps, irrespective whether they are integrable or not, which have a number of invariants.\\

As an illustration, let
\begin{equation}
H(x,y)=y{1+bx\over 1+cx}
,\qquad b,\ c:\ {\rm constants}
\label{H(x,y)=y(1-bx)}
\end{equation}
be the invariant and try to find a two dimensional map which reduces to
\begin{equation}
x\rightarrow X=h{x+a\over 1+bx}(1+cx).
\label{xrightarrow X=hx+aover 1+bx(1+cx)}
\end{equation}
where $h$ is the value of $H(x,y)$. Following to our prescription we immediately find the map
\begin{equation}
(x,y)\rightarrow (X,Y)=\left((x+a)y,\ y{(1+bx)(1+cy(x+a))\over (1+cx)(1+by(x+a))}\right).
\label{moebiuslogistic 2dim}
\end{equation}

The map (\ref{xrightarrow X=hx+aover 1+bx(1+cx)}) is integrable in the limit $c\rightarrow 0$, since it becomes the M\"obius map. When $c$ is finite the map (\ref{xrightarrow X=hx+aover 1+bx(1+cx)}) is nonintegrable since it is equivalent to the logistic map. Therefore the map (\ref{moebiuslogistic 2dim}) provides an example of nonintegrable two dimensional map which has one invariant. In general the existence of a large number of invariants does not guarantee integrability. In fact we could derive a map which reduces to the logistic map, if we choose $f_1(x)=hx(1-x)$ in (\ref{X_i=f_i}) when there are given $d-1$ invariants.

%%%%%%%%%%%%%%%%%%%%%%%%%%%%%%%%%%%%%%%%%%%%%%
\subsection{Iteration of the M\"obius map}

In the previous subsection we learned that there are many higher dimensional integrable maps which reduce to a common integrable lower dimensional map. We study in this subsection the periodicity conditions for the higher dimensional maps which reduce to the M\"obius map.

Now suppose that a $d$ dimensional map of $\mathbf{x}$ is reduced to the M\"obius map
\begin{equation}
x\rightarrow X=h{x+a\over 1+bx},
\label{Moebius}
\end{equation}
after the elimination of $d-1$ variables. If the map is the one of (\ref{moebiuslogistic 2dim}) with $c=0$, the parameter $h$ is the invariant $y(1+bx)$, whereas $a,b$ are some constants. Generally $a,b$ and $h$ are functions of the invariants, but not dependent on $\mathbf{x}$ otherwise.

The iteration of this map does not change the form of the map but only changes the parameters $a,b,h$. If we write 
\begin{equation}
X^{(n)}=h^{(n)}{x+a^{(n)}\over 1+b^{(n)}x}
\label{mobius}
\end{equation}
after $n$ steps, the $(n+1)$th parameters are related to the $n$th ones by
$$
a^{(n+1)}={a+a^{(n)}h^{(n)}\over h^{(n)}+ab^{(n)}},\quad 
b^{(n+1)}={b^{(n)}+bh^{(n)}\over 1+bh^{(n)}a^{(n)}},\quad 
h^{(n+1)}=h{h^{(n)}+ab^{(n)}\over 1+bh^{(n)}a^{(n)}}.
$$
from which we can determine all parameters iteratively as functions of the initial values $(a,b,h)$.

The periodicity conditions of period $n$ for the map (\ref{Moebius}) are now satisfied if the parameters $(a,b,h)$ satisfy
\begin{equation}
(a^{(n+1)}, b^{(n+1)}, h^{(n+1)})=(a,b,h).
\label{(a_n+1, b_n+1, h_n+1)=(a,b,h)}
\end{equation}
From our argument it is clear that the periodicity conditions do not fix the values of the variable $x$ but impose some constranits on the parameters, hence on the invariants.

Solving (\ref{(a_n+1, b_n+1, h_n+1)=(a,b,h)}) iteratively we find the invariant varieties of periodic points as follows
\begin{eqnarray}
v^{(2)}&=& \{\mathbf{x}\ |\ 1+h=0\}\nonumber\\
v^{(3)}&=&\left\{\mathbf{x}\ \Big|\ {1-h^3\over 1-h}+abh=0\right\}\nonumber\\
v^{(4)}&=& \{\mathbf{x}\ |\ 1+h^2+2abh)=0\}\nonumber\\
v^{(5)}&=& \left\{\mathbf{x}\ \Big|\ {1-h^5\over 1-h}+abh(3+(4+ab)h+3h^2)=0\right\}\nonumber\\
v^{(6)}&=& \{\mathbf{x}\ |\ 1-h+h^2+3abh=0\}\nonumber\\
&\vdots&
\label{ivpp of moebius}
\end{eqnarray}

%%%%%%%%%%%%%%%%%%%%%%%%%%%%%%%%%%%%%%%%%%
\subsection{Biquadratic maps}

We have seen that there exist infinitely many $d$ dimensional integrable maps which reduce to the M\"obius map (\ref{Moebius}). We notice that the M\"obius map (\ref{Moebius}) can be written in a bilinear form $bXx+X-hx-ah=0$. If we study some of well known integrable maps with $d-1$ invariants, however, many of them do not reduce to the M\"obius map but to the one specified by the equation
\begin{equation}
S(X,x;\mathbf{q})=0
\label{S(X,x)=0}
\end{equation}
where
\begin{equation}
S(X,x;\mathbf{q}):=aX^2x^2+b(X+x)Xx+c(X-x)^2+dXx+e(X+x)+f,
\label{S(X,x)}
\end{equation}
and 
\begin{equation}
\mathbf{q}=(a,b,c,d,e,f)\in \mathbf{C}^6
\label{q}
\end{equation}
are parameters dependent of the invariants. We shall call a map $x\rightarrow X$ of this form a `biquadratic map' in the following. Some examples which reduce to the biquadratic map (\ref{S(X,x)=0}) are the symmetric version of the QRT map \cite{QRT, QRT2}, the 3d Lotka-Volterra map of (\ref{3LV}) and the $q$-Painlev\'e IV map (\ref{PIV}) with $\alpha_1=\alpha_2=\alpha_3=1$, as we discuss later.\\

One might wonder that the function (\ref{S(X,x)}) does not determine the image of the map uniquely. Moreover the number of images increases rapidly as we repeat the map. This, however, does not happen because of the symmetry of the map under the exchange of $x$ and $X$. In fact we see that one of the solutions of $S(X,x;\mathbf{q})=0$ is $X=X^{(1)}$ and another is $X=X^{(-1)}$ corresponding to the forward and the backward map of $x$. As we repeat the map we obtain the biquadratic map every time, but with different parameters, as we explain now.

The problem of finding the image $Q(x)$ of the second iteration of the map (\ref{S(X,x)=0}) is equivalent to finding an elimination ideal generated by $S(Q,X;\mathbf{q})$ and $S(X,x;\mathbf{q})$. After the elimination of $X$ the function, say $W^{(2)}(Q,x)$, which generates the 1st elimination ideal, becomes quartic in both $Q$ and $x$. Thus we obtain four solutions of $Q$ for given $x$. Those we expect are the solutions of $S(Q,X^{(\pm 1)};\mathbf{q})=0$, corresponding to the paths $x\rightarrow X^{(\pm 1)}\rightarrow X^{(\pm 2)}$ and $x\rightarrow X^{(\pm 1)}\rightarrow x$, since the $X$, which was eliminated, could be either $X^{(1)}$ or $X^{(-1)}$.

In fact, after some manipulation, we obtain
\begin{equation}
W^{(2)}(Q,x)=(Q-x)^2S(Q,x;\mathbf{q}_2),
\label{(Q-x)^2S_2(Q,x)=0}
\end{equation}
with the new parameters $\mathbf{q}^{(2)}=(a^{(2)},b^{(2)},c^{(2)},d^{(2)},e^{(2)},f^{(2)})$ being given by
\begin{eqnarray}
a^{(2)}&:=& (ae-cb)^2-(ad-2ac-b^2)(be-cd+2c^2),\nonumber\\
b^{(2)}&:=& (ae-cb)(2af-be+cd-4c^2)-(ad-2ac-b^2)(bf-ce),\nonumber\\
c^{(2)}&:=&(af-c^2)^2-(ae-bc)(bf-ce),\label{a_2,...,f_2}\\
d^{(2)}&:=&4(af-c^2)^2-2(ae-bc)(bf-ce)-(be-cd+2c^2)^2\nonumber\\
&&
\quad-(ad-2ac-b^2)(df-2cf-e^2),\nonumber\\
e^{(2)}&:=& (fb-ce)(2af-be+cd-4c^2)-(fd-2fc-e^2)(ea-cb),\nonumber\\
f^{(2)}&:=& (fb-ce)^2-(fd-2fc-e^2)(be-cd+2c^2).\nonumber
\end{eqnarray}

Now we suppose that the $k$th images $X^{(\pm k)}$ of the map (\ref{S(X,x)=0}) are found up to $k=n$ by solving $S(Q,x;\mathbf{q}^{(k)})=0,\ k=2,3,...,n$ for $Q$. 
The $(n+1)$th image $X^{(n+1)}$ of $x$ must be obtained by solving the equation $W^{(n+1)}(Q,x)=0$ which is derived from $S(Q,X;\mathbf{q}^{(n)})=0$ and $S(X,x;\mathbf{q})=0$ after the elimination of $X$. Now by solving $S(Q,X^{(\pm 1)};\mathbf{q}^{(n)})=0$ for $Q$ we obtain $X^{(n\pm 1)}(x)$ and $X^{(-n\pm 1)}(x)$. We know already $X^{(\pm(n-1))}(x)$ as the solutions of $S(Q,x;\mathbf{q}^{(n-1)})=0$. Therefore $W^{(n+1)}(Q,x)$ must have $S(Q,x;\mathbf{q}^{(n-1)})$ as a factor. Since $W^{(n+1)}(Q,x)$ is a biquartic function and $S(Q,x;\mathbf{q}^{(n-1)})$ is a quadratic function of $Q$ and $x$, both are symmetric under the exchange of the variables, another factor of $W^{(n+1)}(Q,x)$ must be a symmetric and biquadratic function of $Q$ and $x$. Such a function can be written uniquely in the form of (\ref{S(X,x)}) itself. Therefore we have
\begin{equation}
W^{(n+1)}(Q,x)=S(Q,x;\mathbf{q}^{(n-1)})S(Q,x;\mathbf{q}^{(n+1)}).
\label{W_n+1}
\end{equation}
The new coefficients $\mathbf{q}^{(n+1)}$ will be read off by comparing both sides of (\ref{W_n+1}) as functions of $\mathbf{q}$ and $\mathbf{q}^{(n)}$. Using the notation $(g_\wedge g')_n=g{g'}^{(n)}-g'g^{(n)}$, the results are as follows:
\begin{eqnarray}
a^{(n+1)}&=&{1\over a^{(n-1)}}\Big((a_\wedge c)_n^2-(a_\wedge b)_n(b_\wedge c)_n\Big),\nonumber\\
b^{(n+1)}&=&{1\over a^{(n-1)}}\Bigg({b^{(n-1)}\over a^{(n-1)}}\Big((a_\wedge b)_n(b_\wedge c)_n-(a_\wedge c)_n^2\Big)
+(a_\wedge c)_n\Big((a_\wedge e)_n+2(b_\wedge c)_n\Big)\nonumber\\
&&\qquad-\ {1\over 2}\Big((a_\wedge b)_n(b_\wedge e)_n-(a_\wedge b)_n(c_\wedge d)_n+(a_\wedge d)_n(b_\wedge c)_n\Big)\Bigg),\nonumber\\
c^{(n+1)}&=&{1\over 2c^{(n-1)}}\Bigg((ce^{(n)}-bf^{(n)})(ae^{(n)}-bc^{(n)})+(cb^{(n)}-ea^{(n)})(fb^{(n)}-ec^{(n)})\nonumber\\
&&
\qquad\qquad +(af^{(n)}-cc^{(n)})^2+(fa^{(n)}-cc^{(n)})^2\Bigg),\nonumber\\
d^{(n+1)} &=& {1\over d^{(n-1)}}\Bigg(-f^{(n-1)}a^{(n+1)}-a^{(n-1)}f^{(n+1)}-4b^{(n-1)}e^{(n+1)}-4e^{(n-1)}b^{(n+1)}+(a_\wedge f)_n^2\nonumber\\
&+&(c_\wedge d)_n^2-(a_\wedge b)_n(e_\wedge f )_n-(b_\wedge c)_n(c_\wedge e)_n+(a_\wedge d)_n(d_\wedge f)_n+2(b_\wedge e)_n(a_\wedge f)_n\nonumber\\
&-&
\Big(3(c_\wedge e)_n-(b_\wedge f)_n-(d_\wedge e)_n\Big)\Big(3(b_\wedge c)_n-(a_\wedge e)_n-(b_\wedge d)_n\Big)\nonumber\\
&&\qquad\qquad +2\Big((a_\wedge d)_n-(a_\wedge c)_n\Big)\Big((c_\wedge f)_n-(d_\wedge f)_n\Big)\nonumber\\
&&\qquad\qquad +2\Big((b_\wedge c)_n+(a_\wedge e)_n\Big)\Big((b_\wedge f)_n+(c_\wedge e)_n\Big)\Bigg),
\nonumber\\
e^{(n+1)}&=&{1\over f^{(n-1)}}\Bigg({e^{(n-1)}\over f^{(n-1)}}\Big((f_\wedge e)_n(e_\wedge c)_n-(f_\wedge c)_n^2\Big)
+(f_\wedge c)_n\Big((f_\wedge b)_n+2(e_\wedge c)_n\Big)\nonumber\\
&&\qquad\qquad-\ {1\over 2}\Big((f_\wedge e)_n(e_\wedge b)_n-(f_\wedge e)_n(c_\wedge d)_n+(f_\wedge d)_n(e_\wedge c)_n\Big)\Bigg),\nonumber\\
f^{(n+1)}&=&
{1\over f^{(n-1)}}\Big((f_\wedge c)_n^2-(f_\wedge e)_n(e_\wedge c)_n\Big).\label{n+1th parameters}
\end{eqnarray}
%%%%%%%%%%%%%%

\vglue0.5cm

Our problem of studying the behaviour of the map (\ref{S(X,x)=0}) has been converted to studying the behaviour of the parameters $\mathbf{q}^{(n)}=(a^{(n)},b^{(n)},...,f^{(n)})$. These parameters determine a trajectory of iteration of the map (\ref{S(X,x)=0}):
$$
\cdots \longleftarrow X^{(-2)}\longleftarrow X^{(-1)}\longleftarrow x\longrightarrow X^{(1)}\longrightarrow X^{(2)}\longrightarrow \cdots .
$$
Namely we can consider (\ref{n+1th parameters}) as a map $\mathbf{q}^{(n)}\ \rightarrow\ \mathbf{q}^{(n+1)}$ in $\mathbf{C}^6$.\\

The periodicity conditions of period $n$ for the map $x\rightarrow X$ defined by (\ref{S(X,x)=0}) are satisfied if the parameters satisfy
\begin{equation}
\mathbf{q}^{(n+1)}(\mathbf{q})=\mathbf{q}.
\label{p_n=p_1}
\end{equation}
Despite the complicated expression of the relation (\ref{n+1th parameters}), we observe a special dependence on the $n$th parameters $\mathbf{q}^{(n)}$. Besides $c^{(n+1)}$, the dependence of the $(n+1)$th parameters on the $n$th ones is always in the form $(g_\wedge g')_n=g{g'}^{(n)}-g'g^{(n)}$. They all vanish simultaneously when the periodicity conditions (\ref{p_n=p_1}) are `fully correlated'. In other words if there exists a function $\gamma^{(n+1)}(\mathbf{q})$ such that
\begin{eqnarray}
\mathbf{q}^{(n)}(\mathbf{q})&=&\mathbf{q}+\gamma^{(n+1)}(\mathbf{q})\hat{\mathbf{q}}^{(n)}(\mathbf{q})
\label{a_n=a+gamma a}
\end{eqnarray}
so that (\ref{p_n=p_1}) is satisfied by a single condition $\gamma^{n}(\mathbf{q})=0$. Here we use the same notation $\gamma^{(n)}$, which we used already as a function of $h$ in higher dimension. Although they arise in different contexts they turn out to be the same object as we will see later.

When (\ref{a_n=a+gamma a}) holds, the equation $S(Q,x;\mathbf{q}^{(n+1)})=0$ can be written as
\begin{equation}
c^{(n+1)}(Q-x)^2+\gamma^{(n+1)}(\mathbf{q})K^{(n+1)}(Q,x)=0.
\label{c_n(Q-x)^2+gamma_nK_n(Q,x)=0}
\end{equation}
Here 
\begin{eqnarray*}
K^{(n+1)}(Q,x)&=&\hat a^{(n+1)}Q^2x^2+\hat b^{(n+1)}(Q+x)Qx+\hat d^{(n+1)}Qx+\hat e^{(n+1)}(Q+x)+\hat f^{(n+1)},
\label{K_n}
\end{eqnarray*}
and $\hat a^{(n+1)}$, for instance, is obtained from $a^{(n+1)}$ simply replacing $(g_\wedge g')_n$ by $(\hat g_\wedge{\hat g}')_n$. If $Q$ is a point of period $n+1$, the first term of (\ref{c_n(Q-x)^2+gamma_nK_n(Q,x)=0}) vanishes. Hence the periodicity condition requires for the second term to vanish. This is certainly satisfied for arbitrary $x$ if $\gamma^{(n+1)}(\mathbf{q})=0$, namely when the periodicity conditions for the parameters $\mathbf{q}^{(n)}$ are fully correlated. The other possible solutions obtained by solving $K^{(n+1)}(x,x)=0$ will not correspond to the points of period $n+1$, but represent the fixed points or the points of periods which divide $n+1$.\\

Let us present the functions $\gamma^{(n)}(\mathbf{q})$ explicitly in the cases of small number of $n$. We can show
\begin{eqnarray*}
(a_\wedge b)_2&=&(af-eb-3c^2+cd)(2a^2e-abd+b^3)\\
(a_\wedge c)_2&=&(af-eb-3c^2+cd)(a^2f+ac^2-acd+b^2c)\\
(b_\wedge c)_2&=&(af-eb-3c^2+cd)(2ace-abf-bc^2)\\
&\vdots&\\
(e_\wedge f)_2&=&(af-eb-3c^2+cd)(edf-e^3-2bf^2),
\end{eqnarray*}
from which we find $\gamma^{(3)}(\mathbf{q})$. We can derive other results iteratively as follows:
\begin{eqnarray}
\gamma^{(3)}(\mathbf{q})&=&af-be-3c^2+cd,\nonumber\\
\gamma^{(4)}(\mathbf{q})&=&2acf-adf+b^2f+ae^2-2c^3+c^2d-2bce,
\nonumber\\
\gamma^{(5)}(\mathbf{q})&=&
a^3f^3+\Big(-cf^2d+2cfe^2+fde^2-3ebf^2-e^4-c^2f^2\Big)a^2\nonumber\\
&+&\Big(-13c^4f+18c^3fd+de^3b+2cf^2b^2+7dc^2e^2-ce^2d^2-2ce^3b\nonumber\\
&+&2c^2feb-7fd^2c^2-14c^3e^2+cd^3f+fb^2e^2+f^2db^2-ebd^2f\Big)a\nonumber\\
&-&
cd^2b^2f-b^3e^3-4c^3deb+cdb^2e^2+13ec^4b-f^2b^4+7fb^2c^2d\nonumber\\
&+&
c^4d^2-5c^5d+5c^6-2fb^3ec-e^2c^2b^2+eb^3df-14fb^2c^3,
\label{gamma_3(p)}
\end{eqnarray}
and so on. The formula (\ref{n+1th parameters}) enables us to continue finding a series of $\gamma^{(n)}(\mathbf{q})$ systematically. From the way of this construction it is apparent that all $\gamma$'s are functions of the invariants alone if one of them is so, a fact being consistent with our theorem.

%%%%%%%%%%%%%%%%%%%%%%%%%%%%%%

\subsection{Maps reduced to the biquadratic one}

Having established the scheme of generating the invariant varieties of periodic points of the map (\ref{S(X,x)=0}), we are going to present some examples of higher dimensional maps which reduce to the biquadratic one by using invariants. If $\{X_j=F_j(\mathbf{x}) ,\  j=1,2,...,d\}$ is the map and $H_i(\mathbf{x}),\ i=1,2,...,d-1$ are the invariants, we calculate the $2(d-1)$th elimination ideal of the set of the functions
$$
\{ X_j-F_j(\mathbf{x}),\ H_i(\mathbf{x})-h_i,\ j=1,2,...,d,\ i=1,2,...,d-1 \}
$$
by eliminating the $2(d-1)$ variables $(x_2,x_3,...,x_d,X_2,X_3,...,X_d)$.

%%%%%%%%%%%%%%%%%%%%%%%%%%%%%%%%%
\subsubsection{The symmetric QRT map}

Consider the two dimensional map
\begin{equation}
(x,y)\rightarrow (X,Y)=\left(y,\ {\eta'(y)\rho''(y)-\rho'(y)\eta''(y)-x\Big(\rho'(y)\phi''(y)-\phi'(y)\rho''(y)\Big)\over
\rho'(y)\phi''(y)-\phi'(y)\rho''(y)-x\Big(\phi'(y)\eta''(y)-\eta'(y)\phi''(y)\Big)}\right).
\label{QRT 2d map}
\end{equation}
Here
\begin{eqnarray*}
\phi'(x):=a'x^2+b'x+c',\qquad\quad&\quad& \phi''(x):=a''x^2+b''x+c'',\\
\eta'(x):=b'x^2+(d'-2c')x+e',&\quad& \eta''(x):=b''x^2+(d''-2c'')x+e'',\\
\rho'(x):=c'x^2+e'x+f',\qquad\quad&\quad& \rho''(x):=c''x^2+e''x+f'',
\end{eqnarray*}
and  $\mathbf{q}'=(a',b',c',d',e',f')$ and $\mathbf{q}''=(a'',b'',c'',d'',e'',f'')$ are constants. If we write $(Y,y,x)$ as $(x^{(n+1)},x^{(n)},x^{(n-1)})$, this is nothing but the symmetric case of the well known QRT equation \cite{QRT, QRT2}
\begin{equation}
x^{(n+1)}={\eta'(x^{(n)})\rho''(x^{(n)})-\rho'(x^{(n)})\eta''(x^{(n)})-x^{(n-1)}\Big(\rho'(x^{(n)})\phi''(x^{(n)})-\phi'(x^{(n)})\rho''(x^{(n)})\Big)\over
\rho'(x^{(n)})\phi''(x^{(n)})-\phi'(x^{(n)})\rho''(x^{(n)})-x^{(n-1)}\Big(\phi'(x^{(n)})\eta''(x^{(n)})-\eta'(x^{(n)})\phi''(x^{(n)})\Big)}.
\label{QRT equation}
\end{equation}
The map (\ref{QRT 2d map}) has an invariant
\begin{equation}
H(x,y)=-\ {\phi'(x)y^2+\eta'(x)y+\rho'(x)\over \phi''(x)y^2+\eta''(x)y+\rho''(x)},
\label{h QRT}
\end{equation}
hence it can be reduced to one dimensional map $x\rightarrow X$. The calculation of the 2nd elimination ideal is rather trivial in this case. If $y$ and $Y$ are eliminated by using the invariant $H(x,y)=h$, the result we obtain is
\begin{equation}
\phi(x)X^2+\eta(x)X+\rho(x)=0
\label{phi(x)X^2+eta(x)X+rho(x)=0}
\end{equation}
where
$$
\phi(x):=ax^2+bx+c,\quad \eta(x):=bx^2+(d-2c)x+e,\quad
\rho(x):=cx^2+ex+f,
$$
with 
$$
\mathbf{q}=\mathbf{q}'+h\mathbf{q}''.
$$
If we identify $\mathbf{q}=(a,b,c,d,e,f)$ with the one of (\ref{q}), the map (\ref{phi(x)X^2+eta(x)X+rho(x)=0}) is exactly the biquadratic map (\ref{S(X,x)=0}). 

In the theory of the QRT map the formula (\ref{phi(x)X^2+eta(x)X+rho(x)=0}) is called an invariant curve. The problem of solving the equation (\ref{QRT 2d map}) is now converted to finding the coefficients of (\ref{phi(x)X^2+eta(x)X+rho(x)=0}) iteratively. Our general formula (\ref{gamma_3(p)}) gives us the explicit expressions of the invariant varieties of the symmetric QRT map (\ref{QRT 2d map}).

Recently Tsuda \cite{Tsuda} discussed the QRT map from a geometrical viewpoint and obtained many interesting results. In particular all periodic maps of QRT were classified corresponding to particular combinations of the parameters. 
%%%%%%%%%%%%%%%%%%%%%%%%%%%%%%%%%%%%%%%%%%%%%

\subsubsection{3d Lotka-Volterra map}

After the elimination of $x_2, x_3$ and $X_2, X_3$ from (\ref{3LV}) by using (\ref{r,s}) and denoting $x_1=x$ we see that the 4th elimination ideal is given by the biquadratic function (\ref{S(X,x)}) with the coefficients
\begin{eqnarray*}
&a=r+1,\quad b=s-2r-1,\quad c=r-s,&\nonumber\\
&d=s^2+rs+5r-2s+1,\quad e=-r(s+1),\quad f=0.&
\end{eqnarray*}

From the general argument we obtain the result of first iteration simply substituting these data into (\ref{a_2,...,f_2}). They are given by
\begin{eqnarray*}
a^{(2)}&=&(s+1)^2s(r^2-rs^2-s-3rs),\nonumber\\
b^{(2)}&=&(s+1)^2s(2r^2s+s+5rs-2r^2-r^3-s^2),\nonumber\\
c^{(2)}&=&(s-r)s(r+1)(s^2-r^2s-3rs-r)\label{LV p_2},\\
d^{(2)}&=&(s+1)^2s(2rs^2+2s^2-3r^2s-8rs-s+r^3s+r^4+5r^3+6r^2-s^3),\nonumber\\
e^{(2)}&=&(s+1)^2rs(s^2-rs+s-2r-r^3-2r^2),\nonumber\\
f^{(2)}&=&(s+1)^2r^2s(r^2-rs+r+s^2+s+1).\nonumber
\end{eqnarray*}
Notice that all parameters apart from $c^{(2)}$ are proportional to a common factor $(s+1)^2$. This is a result which we cannot derive generally from (\ref{a_2,...,f_2}) since they are not factorized. On the other hand this factor is exactly the one we expect from our previous result (\ref{period 2}), although this is a special behaviour of the 3d Lotka-Volterra map.

The periodicity conditions of period 3 and higher can be derived similarly and the associated functions $\gamma^{(3)}$, $\gamma^{(4)}$ and $\gamma^{(5)}(\mathbf{q})$ can be read off from (\ref{gamma_3(p)}) directly. We find
\begin{eqnarray*}
\gamma^{(3)}(r,s)&=&r^2+s^2-rs+r+s+1,\\
\gamma^{(4)}(r,s)&=&3rs+s+s^3-3s^2r+r^3s+6r^2s-r^3,\\
\gamma^{(5)}(r,s)&=&r^3s^4-r^3s^2-6r^4s^5+10r^3s^6+3s^5r+s^6+s^5+3r^4s^4-3r^5s^3\\
&&
-6r^4s^3-r^6s^3+3r^5s^4+s^4+21s^4r^2+6s^4r+r^3s^7+s^7+27s^5r^2\\
&&
-3s^6r-r^3s^5+21r^2s^6-10r^3s^3-6rs^7+s^8.
\end{eqnarray*}
These include again precisely the conditions expected from our direct calculations (\ref{period 3}), (\ref{period 4}) and (\ref{period 5}). 

It will be worthwhile to mention about the uniqueness of the reduction. If we had reduced the map (\ref{3LV}) by eliminating $x_2$ and $x_3$ one by one in different order we should have obtained two different maps. They correspond to the exchange of $X^{(1)}$ and $X^{(-1)}$. This difference becomes irrelevant in our procedure since the 4th elimination ideal (\ref{S(X,x)}) combines them together.

The 4d Lotka-Volterra map (\ref{4dLV}) also satisfies $d-p=1$. This map, however, does not reduce to the biquadratic map, but to a biquartic one.

%%%%%%%%%%%%%%%%%%%%%%%%%
\subsubsection{Painlev\'e IV map}

The $q$-Painlev\'e map (\ref{PIV}) can be converted into the biquadratic map (\ref{S(X,x)=0}) when the parameters are fixed at $\alpha_1=\alpha_2=\alpha_3=1$. In fact, after the elimination of $x_2$ and $x_3$ we find
\begin{eqnarray*}
&a=s+1,\qquad b=-r-1,\qquad c=0,&\\
&d=r^2+4r+sr-s+1,\qquad e=-r^2-r,\qquad f=r^2-rs,&
\end{eqnarray*}
which reproduce (\ref{PIVinvariant varieties}).

The $q$-Painlev\'e V map (\ref{Painleve V}) at $\alpha_1=\alpha_2=\alpha_3=\alpha_4=1$ does not reduce to the biquadratic map but to a biquartic map.\\

%%%%%%%%%%%%%%%%%%%%%%%%%%%%%%%%%%%%%%%%%%%%%%%%%
%%%%%%%%%%%%%%%%%%%%%%%%%%%%%%%%%%%
\section{Transition between Integrable and Nonintegrable Maps}

We have not discussed, so far, the behaviour of periodic points of nonintegrable maps. The purpose of this section is to clarify the transition between integrable and nonintegrable maps. In particular we are interested in how a set of isolated periodic points in one side turns to an invariant variety in other side. 

It will be useful if there is a simple model which interpolates the integrable and nonintegrable maps, such that we can study explicitly how the transition of the periodic points takes place between the two maps. The method developed in \S 4.1 provides us many such maps, in the sense that the study of one low dimensional map is sufficient to know the behavior of many higher dimensional maps asssociated to it, irrespective whether the map is integrable or not.

To be specific we study, in this section, the map (\ref{xrightarrow X=hx+aover 1+bx(1+cx)}) which is integrable only in the limit $c\rightarrow 0$. All information of higher dimensions are included in the parameters $a,b,c$ and $h$. But we assume $c$ is a constant parameter so that we can control the value of $c$ by hand without changing the dependence of the higher dimensional coordinates. Moreover it is convenient to introduce the new variable $z$ and convert the map (\ref{xrightarrow X=hx+aover 1+bx(1+cx)}) to the normal form of the one dimensional rational maps of degree 2
\begin{equation}
z\ \ \rightarrow\ \  Z=z{\lambda' +z\over 1+\lambda z},
\label{Z}
\end{equation}
which has been studied intensively in the literature \cite{FM, FM2}. The correspondence between $x$ and the new variable $z$
is given by
$$
x={\Big((1-h+2hab-hac)q_--q_+^2\Big)z+(1-h+2hab-hac)q_-+q_+^2\over \big((2hc-abch-b-bh)q_-+bq_+^2\big)z+(2hc-abch-b-bh)q_--bq_+^2},
$$
and the new parameters are related by
\begin{eqnarray}
\lambda&=&{(2ch-b)q_+^2-(2hc-abch-b-bh)q_-
\over
bq_+^2-(2hc-abch-b-bh)q_-}
\label{lambda}\\
\lambda'&=&
{(2ch-b)q_+^2+(2hc-abch-b-bh)q_-
\over
bq_+^2+(2hc-abch-b-bh)q_-}
\label{lambda'}
\end{eqnarray}
where
$$
q_\pm^2=(1-h-hac)^2\pm 4ah(b-hc).
$$
The map (\ref{Z}) is integrable when $\lambda\lambda'=1$, hence $c=0$, but not integrable otherwise.

Repeating the map (\ref{Z}) $n$ times the image of $z$ can be written as
\begin{eqnarray}
Z^{(n)}&=&{1\over \lambda}Z^{(n-1)}{\lambda'+Z^{(n-1)}\over \lambda^{-1}+Z^{(n-1)}}
\label{Z^(n)1}\\
&=&
{1\over \lambda^n}z\left({\lambda'+z\over \lambda^{-1}+z}\right)\left({\lambda'+Z^{(1)}\over \lambda^{-1}+Z^{(1)}}\right)\left({\lambda'+Z^{(2)}\over \lambda^{-1}+Z^{(2)}}\right)\cdots\left({\lambda'+Z^{(n-1)}\over \lambda^{-1}+Z^{(n-1)}}\right).
\label{Z^(n)2}
\end{eqnarray}
If $k_{n-1}$ is the degree of the rational polynomial $Z^{(n-1)}$, we see from (\ref{Z^(n)1}) that $k_n=2k_{n-1}$, hence $k_n=2^n$. Periodic points are obtained by solving $Z^{(n)}=z$. When $n$ is a prime number, the number $\#_n$ of the periodic points of period $n$ is $k_n-2=2^n-2$. Here subtraction of 2 corresponds to two fixed points. If $n$ is not prime the contribution from the divisors must be also subtracted. In this way we find that the number increases as fast as 
$$
(\#_2,\#_3,\#_4,\cdots)=(2,6,12,30,48,126,240,504,\cdots).
$$

In the integrable limit $\lambda\lambda'=1$, this increase ceases owing to the cancellation of factors in the numerator and the denominator as seen in the second expression (\ref{Z^(n)2}), and the map becomes
\begin{equation}
Z^{(n)}=\lambda^{-n}z.
\label{Z=h^-nz}
\end{equation}
It is apparent that there is no periodic point unless $\lambda$ satisfies $\lambda^n=1$ or
$$
\lambda=e^{i 2\pi k/n},\qquad k=1,2,...,n. 
$$
We now recall that $\lambda$ is a function of the invariants of a higher dimensional map. Let us write the right hand side of (\ref{lambda}) as $\lambda(h,a,b,c)$. Then the invariant varieties of periodic points of the map are given by
\begin{equation}
v^{(n)}=\{ \mathbf{x}\ |\ \lambda(h,a,b,0)^n-1=0\},\quad n=2,3,4,....
\label{varieties of 2d map}
\end{equation}
which provides another expression of (\ref{ivpp of moebius}).
\\

All isolated periodic points suddenly disappear in the integrable limit, and the invariant varieties (\ref{varieties of 2d map}) appear. In order to uncover this trick of the transition we must know where the isolated periodic points were right before they disappeared. After some manipulation we find that the right hand side of (\ref{Z^(n)2}) admits the following expression 
$$
Z^{(n)}=
{z\over \lambda^n}\ {\displaystyle{\prod_{k=-1}^{n-2}(z+\lambda^k)^{2^{n-k-2}}+(\lambda\lambda'-1)P_n}\over
\displaystyle{\prod_{k=-1}^{n-2}(z+\lambda^k)^{2^{n-k-2}}+(\lambda\lambda'-1)Q_n}}
$$
where $P_n$ and $Q_n$ are polynomials of $z$. The periodicity condition $Z^{(n)}=z$ requires
\begin{equation}
(\lambda^n-1)\prod_{k=-1}^{n-2}(z+\lambda^k)^{2^{n-k-2}}
=
(\lambda\lambda'-1)(P_n-\lambda^nQ_n)
\label{periodicity cond for small}
\end{equation}
to hold. 

When $\lambda\lambda'=1$ we obtain solutions of (\ref{periodicity cond for small}) at
\begin{equation}
z=-1/\lambda,-1,-\lambda,\cdots,-\lambda^{n-2}
\label{-1/h,-1,-h..}
\end{equation}
if $\lambda^n\ne 1$. These points are, however, certainly not the periodic points on the invariant varieties (\ref{varieties of 2d map}). Suppose $\lambda\lambda'-1$ is very small but not zero. Then the periodicity condition (\ref{periodicity cond for small}) is satisfied {\it iff} $\mathbf{x}$ is very close to one of the points satisfying (\ref{-1/h,-1,-h..}). Therefore the periodic points were in the neighbourhood of these points right before $\lambda\lambda'$ reached 1.\\

Finally we consider the connection of our result with the study of the Julia set of the map (\ref{Z}). The Julia set can be obtained by calculating iteratively the inverse map of (\ref{Z}) starting from one of repulsive periodic points. Since the inverse map of (\ref{Z})
\begin{equation}
Z\rightarrow z=\left\{\begin{array}{l}{1\over 2}\big(\lambda Z-\lambda'+\sqrt{(\lambda Z+\lambda')^2+4(1-\lambda\lambda')Z}\big)\\
\\
{1\over 2}\big(\lambda Z-\lambda'-\sqrt{(\lambda Z+\lambda')^2+4(1-\lambda\lambda')Z}\big)
\end{array}
\right.
\end{equation}
has two branches, the number of the points increases rapidly and they form a fractal structure including many sets of periodic points. 

As the map approaches to the integrable limit $\lambda\lambda'\rightarrow 1$, the inverse map simplifies significantly turning to
\begin{equation}
z=\left\{\begin{array}{l}
\lambda Z\\
\\
-{1/\lambda}
\end{array}\right.
\end{equation}
and one of the branches is frozen at the point $z=-1/\lambda$. The degeneracy of many maps to this one particular point causes a collapse of the Julia set. We have shown in the papers \cite{YS, SSS} that the Julia set of the map (\ref{Z}) approaches uniformly to the set $\lim_{n\rightarrow\infty}J_n$, in the limit $\lambda\lambda'\rightarrow 1$, where$$
J_n=\{0,-1/\lambda,-1,-\lambda,\cdots,-\lambda^{n-2}\}.
$$
Therefore the points of (\ref{-1/h,-1,-h..}) are ``fossils'' of the Julia set after the collapse. We learn, from this fact, that the invariant varieties of periodic points are created independent from the Julia set in the integrable limit. The Julia set itself ceases to be a set of periodic points owing to the degeneracy of the critical points at $z=-1/\lambda$.\\

A simpler model which presents similar property that we discussed in this section was invented by S. Onozawa. Using that model he explored more detail of the transition between integrable and nonintegrable maps from a different point of view\cite{Onozawa}. 

%%%%%%%%%%%%%%%%%%%%%%%%%%%%%%%%%%%%%%%
\section{Concluding Remarks}

Throughout this paper we have examined many examples of invariant varieties of periodic points in order to support our conjecture in \S 1. To conclude this paper we would like to summarize some remarks on our study.

\begin{enumerate}
\item
In contrast to the continuous time Hamiltonian flow, the number of invariants does not play the critical role to decide integrability of a map. In fact there are higher dimensional maps which have many invariants but reduce to nonintegrable maps, while some maps, like the q-Painlev\'e maps, have no invariants but are integrable.

\item
If a periodic point of some period is isolated, the periodic map can start only from certain limited points. When the point is on an invariant variety of periodic points, on the other hand, the map can start from arbitrary point of the variety. Therefore the conjecture in \S 1 can be rephrased as ``an existence of movable periodic points, dependent on initial values, indicates integrability of the map.''

\item
The Poincar\'e sections of a periodic orbit of an integrable system produce a smooth curve in a plane of the phase space as initial values are changed continuously. If we consider the intersections as a sequence of a discrete map, we can think of the curve as an invariant variety of periodic points. When the map is perturbed, our lemma requires that all points on the invariant variety are frozen by the Birkhoff's fixed point theorem.

\item
It is well known that some higher dimensional nonintegrable maps are characterized by strange attractors in which a dense set of periodic points of all periods are condenced. As we consider the attractor in the complex space, it is embedded in the Julia set in some cases \cite{BS,SII}. Therefore it is a different object from the invariant variety of periodic points, since the latter is a variety formed by a set of periodic points of a single period.

\item
We have concerned in this paper with a sufficient condition for the integrability of a rational map. The necessary condition for the integrability will be given {\it iff} the precise notion of integrability is known. Our theorem can not discriminate an integrable map from nonintegrable one if the periodicity conditions are uncorrelated in all periods. The study of the $q$-Painlev\'e maps in \S 3-3 shows the difficulty of clarifying the border of these two regimes.

\item
In order to clarify the phenomenon which takes place at the boder between integarable and nonintegrable regimes we have investigated in detail the transition of a map from one side to another in \S 5. It is found that the source of the periodic points are clearly different. Namely the periodicity conditions are imposed on the variables to determine their position if the map is nonintegable, while, if the map is integrable, they are imposed on the invariants to determine their relations among themselves. This argument, however, does not apply to an integrable map which has no invariant. We also would like to mention that there are some subtleties involved in defining integrability even when all periodic points are neutral \cite{Berry, Bogomolny}. 
\end{enumerate}

\noindent
{\large{\bf Acknowledgements}}

The idea of this study arose from discussions with Dr. Katsuhiko Yoshida some years ago. We would like to thank him and also Mr. Show Onozawa for many interesting discussions. We would like to express our special thanks to Prof. Martin Guest for suggesions concerning the exposition. We also thank Prof. Akira Shudo for many valuable comments especially on the topics related to his own works.\\

%%%%%%%%%%%%%%%%%%%%%%%%%%%%%%%%%%%%%%%%%%%%%
\noindent
{\large{\bf Appendix: \ Invariants of the Lotka-Volterra Maps}}\\

Let us define the matrices
$$
R(t)=
\left(\begin{array}{ccccc}
1-X_1^{(t)}&1&0&\cdots&0\cr
0&1-X_2^{(t)}&1&&\vdots\cr
\vdots&&\ddots&&0\cr
0&\cdots&0&1-X_{d-1}^{(t)}&1\cr
1&0&\cdots&0&1-X_d^{(t)}\cr
\end{array}\right),
$$$$
L(t)=
\left(\begin{array}{cccccc}
0&0&\cdots&0&X_d^{(t)}&1\cr
1&0&&&0&X_1^{(t)}\cr
X_2^{(t)}&1&0&\cdots&&0\cr
0&X_3^{(t)}&1&0&\cdots&\cr
\vdots&&&\ddots&&\vdots\cr
0&\cdots&0&X_{d-1}^{(t)}&1&0\cr
\end{array}\right).
$$
A straightforward calculation will show that the Lotka-Volterra equations (\ref{LV eq}) are equivalent to the following matrix formula,
$$
L(t+1)R(t+1)=R(t)L(t).
$$
To find invariants we define
$$
A(t):=L(t)R(t)=
\left(\begin{array}{ccccccc}
1&0&&\cdots&0&p_d&1\cr
1&1&0&&&0&p_1\cr
p_2&1&1&0&\cdots&&0\cr
0&p_3&1&1&\cdots&\cr
\vdots&&&&\ddots&&\vdots\cr
0&&&&1&1&0\cr
0&&\cdots&0&p_{d-1}&1&1\cr
\end{array}\right),
$$
where we used the notation $p_j:=X_j^{(t)}(1-X_{j-1}^{(t)})$. Since
$$
A(t+1)=L(t+1)R(t+1)=R(t)A(t)R^{-1}(t),
$$
eigenvalues of $A$ are invariant. If we write
$$
\det(A-\lambda)=(-1)^{d-1}\sum_{k=0}^dH_{k}(\lambda-1)^k,
$$
the set of coefficients $H_0,H_1,...,H_d$ is also invariant. Comparing both sides we find
$$
H_k=\left\{\begin{array}{cl}
1-(-1)^dp_1p_2\cdots p_d,&\quad k=0\cr
\displaystyle{{\sum}'_{j_1,j_2,...,j_k}p_{j_1}p_{j_2}\cdots p_{j_k},}&\quad k=1,2,...,[d/2]\cr
0,&\quad k=[d/2]+1,...,d-1\cr
-1,&\quad k=d.\cr
\end{array}\right.
%\label{H_k}
$$
Here $[d/2]=d/2$ if $d$ is even and $[d/2]=(d-1)/2$ if $d$ is odd. The prime in the summation $\sum'$ means that the summation must be taken over all possible combinations $j_1,j_2,...,j_k$ but excluding direct neighbours. Since $H_0$ can be represented by other $H_k$'s and
$$
r=X_1^{(t)}X_2^{(t)}\cdots X_d^{(t)},
$$
it is convenient to use $r$ instead of $H_0$.

%%%%%%%%%%%%%%%%%%%%%%%%%%%%%%%%%%%%%%%%%%%%%%%%%%

\end{document}